\documentclass[1p,times]{elsarticle}

\usepackage{ecrc}

\volume{00}

\firstpage{1}

\journalname{Physics Procedia}

\runauth{}

\jid{procs}

\jnltitlelogo{Physics Procedia}

\usepackage{amssymb}

\usepackage[figuresright]{rotating}

\newcommand{\dir}{Figs}

\begin{document}

\begin{frontmatter}



\dochead{}

\title{A method to compute absolute free energies or enthalpies of fluids}


\author[fs]{Friederike Schmid}
\author[ts]{Tanja Schilling}

\address[fs]{Institut f\"ur Physik, Johannes Gutenberg Universit\"at, Staudinger Weg 9, D55099 Mainz, Germany}
\address[ts]{Theory of Soft Condensed Matter, Universit\'e du Luxembourg, 162a, rue de la faiencerie, 
  BRB0.07, 1511 Luxembourg}

\begin{abstract}
We propose a new method to compute the free energy or enthalpy of fluids or
disordered solids by computer simulation . The main idea is to construct a
reference system by freezing one representative configuration, and then carry
out a thermodynamic integration. We present a strategy and an algorithm which
allows to sample the thermodynamic integration path even in the case of
liquids, despite the fact that the particles can diffuse freely through the
system. The method is described in detail and illustrated with applications to
hard sphere fluids and solids with mobile defects.
\end{abstract}


\end{frontmatter}



\section{Introduction}
\label{sec:introduction}

The free energy or free enthalpy is a central quantity in the statistical
physics and thermodynamics of equilibrium systems, and there has traditionally
been large interest in computing free energies in many areas of science
\cite{chipot07}.  Free energies need to be evaluated if one wishes to locate
first order phase boundaries accurately, to compare differently folded
conformations of large molecules, or to characterize adsorption properties. In
all of these cases, the quantities of interest are actually free energy {\em
differences}, and a large number of elegant methods have been devised that
allow to relate the free energies of specific systems with each other
\cite{frenkel02,landau05,earl05,okamoto04,panagiotopoulos00,pablo99,bruce00,wilding00}.
An obvious alternative is to compute the absolute free energies of the two
systems separately.  Much less effort has been devoted to this type of approach.
Computing the absolute free energies of arbitrary disordered or fluid systems such as,
e.g., lipid membranes, solid with highly mobile defects, or nematic liquid
crystals, remains one of the long-standing unsolved problems in computer
simulations.  If a fluid state has a direct connection with an ideal gas state
in phase space, its free energy can be evaluated by standard thermodynamic
integration techniques. In the systems mentioned above, however, the fluid phase
of interest cannot be accessed from the gas phase in an obvious manner without
crossing a first order phase boundary.

Here we propose a method to calculate the free energies of arbitrary liquids
and disordered solids \cite{schilling09}, which will hopefully contribute to solve the
problem. The main idea is to freeze an arbitrary configuration in the phase
of interest -- a 'representative' configuration, obtained, e.g., from a typical
simulation of an equilibrated system -- and to use this frozen 'representative'
configuration as the reference system for a thermodynamic integration. This
idea seems bold and precautions have to be taken to make it work for
fluids. We propose an algorithm that allows to determine the free energies for
simple fluids and solids (as we shall demonstrate for hard spheres), which can
presumably be generalized to complex fluids in a fairly straightforward manner.

To put the method into context, we briefly sketch a few similar proposals in
the past. Our idea is based on the ''Einstein crystal'' method originally
developed by Frenkel and Ladd \cite{frenkel84}, which allows one to calculate the
free energy of a regular crystal structure by establishing a thermodynamic
integration path to a reference system where particles are pinned to the
lattice sites of the same crystal by harmonic springs. This method has been
generalized in various ways. For example, the ''Einstein molecule'' \cite{vega07}
method alleviates the practical difficulties associated with shifts of the
whole crystal during a simulation, and an ''Einstein well'' method was designed
for ''cluster crystals'' where the lattice sites are occupied by several
particles \cite{mladek08}.  Speedy has generalized the Einstein method
for amorphous solids by proposing to use reference systems with irregularly
distributed sites \cite{speedy93}, much in the spirit of our 'representative'
reference system. 

The basic problem with applying the Einstein method to fluids is that
''Einstein'' particles are bound to their reference site -- even in the limit
of very weak springs -- whereas fluid particles explore the whole space. In
infinite space, the transition from the ''Einstein'' system to the free fluid
is thus singular.  In finite space, the singularity disappears, but sampling
the relevant quantities for the thermodynamic integration remains difficult
(see below).  Tyka {\em et al} \cite{tyka07} have recently proposed to remedy
this situation by using a complicated multi-particle spring potential, where
particles are assigned to reference sites such that the {\em total} spring
energy is minimal, under the constraint that there is still a one-to-one
correspondence between particles and reference sites. Thus particles swap
reference sites as they move along. As an application example, Tyka {\em et al}
have calculated the free energies of liquid water and liquid argon. Their
method has some similarities with the method we shall propose below, but our
approach is conceptually simpler, and we do not need to solve a complex
assignment problem for every single configuration.

Our paper is organized as follows. In the next two sections, we present first
the basic strategy of our method, and then the practical implementation.
Extensions of the method, {\em e.g.}, to constant pressure pressure simulations,
are discussed in Section \ref{sec:extensions}. In Section \ref{sec:results}, we
present first results regarding the behavior of the algorithm for different
algorithmic parameters, and applications to hard sphere fluids. We conclude and
summarize in section \ref{sec:summary}.

\section{Basic Strategy}

For clarity, we will constrain the discussion to  monatomic liquids at constant
volume at first.  Extensions, {\em e.g.}, to constant pressure simulations will
be introduced further below (Section \ref{sec:extensions}). Furthermore, we
only consider the configurational part of the free energy, since the kinetic
part can be evaluated trivially \cite{jelitto89}). The configurational energy
is then characterized by a potential function or 'Hamiltonian' 
$U(\{\mathbf{r}_i)\}) + H+0$, which depends on the coordinates $\{ \mathbf{r}_i \}$ of
the particles $i$, and the configurational free energy shall be denoted $F_0$. 

The first step is to choose a 'representative' reference configuration,
$\{\mathbf{r}^{\rm ref}_i\}$, which is typically some arbitrary equilibrated
configuration.  The reference system is defined by the reference Hamiltonian
\begin{equation}
\label{eq:RefHamilton}
H_{\rm ref} (\varepsilon) = \varepsilon\sum_i 
\Phi\left(\frac{|\mathbf{r}_i - \mathbf{r}^{\rm ref}_i|}{r_{\rm cutoff}}\right) 
+ U(\{ \mathbf{r}_i^{\rm ref} \})
\quad , 
\end{equation}
where the function $\Phi(x)$ defines an attractive potential well with $\Phi(x)
<0$ for $x < 1$ and $\Phi \equiv 0$ elsewhere. The last term $U(\{
\mathbf{r}_i^{\rm ref} \})$ denotes the potential energy of hypothetical
particles pinned to the {\em reference} sites and is thus constant.  It is
introduced for convenience.  Note that there is formally a one-to-one
correspondence between particles and reference sites, like in the Einstein
crystal method. However, we want the particles to be indistinguishable (in the
'classical' sense), hence we allow them to swap identities ({\em i.e.}, labels
$i$ and $j$) during the simulation. This indistinguishability together with the
finite range of the well potential (in the Einstein crystal, it is infinite)
helps to avoid the conceptual problems with the transition between the
reference system $H_{\rm ref}$ and the target system $H_0$ mentioned earlier.
Moreover, we shall show below (Section \ref{sec:switch_wells}) that the swap
moves also speed up the equilibration times considerably.  In the reference
system, the particles do not interact directly with each other, hence its free
energy $F_{\rm ref}(\varepsilon)$ can be calculated analytically (see Section
\ref{sec:reference}).

To establish the connection between the reference system $H_{\rm ref}$ and 
the target system $H_0$, we introduce an intermediate model
\begin{equation}
\label{eq:IntHamilton}
H'(\lambda,\varepsilon) = H_{\rm inter}(\lambda) + H_{\rm ref} (\varepsilon)
\end{equation}

that reduces to the target system $H_0$ at $(\lambda = 1, \varepsilon = 0)$ and
to the reference system $H_{\rm ref}(\varepsilon)$ at $\lambda = 0$. For
example, one can choose $H_{\rm inter}(\lambda) = \lambda \: \Delta U$ with
$\Delta U := U(\{\mathbf{r}_i\})- U(\{\mathbf{r}_i^{\rm ref}\})]$.
Alternatively, a more complicated function like $H_{\rm inter}(\lambda) = - k_B
T \ln(1-\lambda+\lambda e^{-\beta \: \Delta U})$ can be more convenient when
dealing with hard core interactions. The free energy of the intermediate system
(\ref{eq:IntHamilton}) shall be denoted $F'(\lambda,\varepsilon)$.  Within the
intermediate model, we can construct a thermodynamic integration path
connecting the reference and target system. In practice, it is often simplest
to evaluate separately the free energy difference associated with switching on
the particle interactions ($\lambda=0 \to 1$) at fixed $\varepsilon$, and the
free energy difference associated with switching off the well potentials in
full presence of particle interactions ($\lambda = 1$). Other integration paths
are also conceivable and may be more effective in some cases.

We will now discuss separately the different practical steps involved in 
this program.

\subsection{Free energy of the reference system}
\label{sec:reference}

We begin with calculating the free energy of the reference system.  The
partition function of a single particle in the volume $V$ subject to a well
potential $\Phi(|\mathbf{r} - \mathbf{r}^{\rm ref}|/r_{\rm cutoff})$ is given by 
\begin{displaymath}
{\cal Q}_1 = V + \int_V {\rm d}\mathbf{r} \: 
\big[ \exp(-\beta \varepsilon \: \Phi) - 1 \big]
=: V + V_0 \: g_{\Phi}(\beta \varepsilon)
\end{displaymath}
with $\beta = 1/k_B T$, where $V_0$ is the volume of a sphere with radius
$r_{\rm cutoff}$. This defines the function
\begin{equation}
\label{eq:g_phi}
g_{\Phi}(a) = {\rm d} \: \int_0^1 {\rm d}x \: x^{{\rm d}-1} \:
[ e^{- a \Phi(x)} - 1]
\end{equation} 
characterizing the effect of a well potential $\Phi(x)$ in 
${\rm d}$ spatial dimensions on the single-particle partition function. 
Below, we shall mostly use a linear well potential,
\begin{equation}
\label{eq:phi_linear}
\Phi(x) = \left\{ \begin{array}{cl}
x-1 \quad & \mbox{for} \: x < 1 \\ 0 \quad & \mbox{otherwise,} 
\end{array} \right.
\end{equation}
giving
\begin{equation}
\label{eq:g_linear}
g_{\rm linear}(a) = 
\frac{\rm d!}{a^{\rm d}} \: \left( e^{a}  \:
- \sum_{k=0}^{\rm d} \frac{a^k}{k!} \right).
\end{equation}
In terms of the characterizing function $g_{\Phi}(a)$, the configurational
free energy of the reference system of $N$ particles subject to the
Hamiltonian $H_{\rm ref}(\varepsilon)$ finally reads
\begin{equation}
\label{eq:ref_FreeEnergy}
  F_{\rm ref}(\varepsilon)
    = N k_B T \left[ \ln\left(\frac{N}{V}\right) 
      - \ln\left(1 +\frac{V_0}{V} g_{\Phi}(\beta \varepsilon)\right)
      \right] + U(\{ \mathbf{r}_i^{\rm ref} \}).
\end{equation}

\subsection{Switching on the particle interactions}
\label{sec:switch_inter}

The next step consists in switching to the intermediate model
$H'(\lambda,\varepsilon)$ and evaluating the free energy difference associated
with taking the interaction parameter $\lambda$ from zero to one. This could be
done by a suitable thermodynamic integration. Here we will describe a different
approach, which is also commonly used in the literature: We carry out a series
of extended ensemble simulations where the parameter $\lambda$ can switch
between two values $\lambda_0$ and $\lambda_1$ ({\em e.g.}, in the simplest
case $\lambda_0=0$ and $\lambda_1=1$) by means of Monte Carlo trial moves,
which are accepted or rejected according to a Metropolis criterion with the
Hamiltonian $H'(\lambda, \varepsilon)$ (Equation (\ref{eq:IntHamilton})). The
free energy difference of two systems with interactions $H_{\rm
inter}(\lambda_0)$ and $H_{\rm inter}(\lambda_1)$ is then given by $\Delta F =
F'(\lambda_1,\varepsilon)-F'(\lambda_0,\varepsilon) = - k_B T
\ln(P_{\lambda_1}/P_{\lambda_0})$, where $P_{\lambda_i}$ is the fraction of
configurations with interaction parameter $\lambda_i$.

The question remains how to choose the steps $\lambda_i$. On the one hand, the
free energy difference $\Delta F$ should not exceed several $k_B T$, otherwise
the system stays in the more favorable state forever. On the other hand, it
should not be too small, otherwise it becomes difficult to control the relative
statistical error. By a simple argument, we can estimate that the 'optimal'
free energy difference should be about 2 $k_B T$:

Consider two systems with a total free energy difference $\Delta F_0$, and we
have the task to evaluate the quantity $q = \exp(- \Delta F_0)$ by extended
ensemble simulations of total length $N$. For simplicity, we assume that every
Monte Carlo sweep results in an independent configuration. We compare two
different simulation strategies: In the first, $q$ is evaluated in one step
within one single simulation run of length $N$, where the two systems are
compared directly. In the second, $q$ is evaluated in $k$ steps, passing
$(k-1)$ intermediate system with free energies equally distributed between the
two systems of interest. This involves $k$ simulation runs of length $N/k$.  In
the first setup (one-step simulation), the fraction of configurations $N_1$ in
the state of higher free energy is given by $N_1/N \sim \langle
N_1/(N-N_1)\rangle = \exp(-\beta \Delta F_0)$ with the relative error $\Delta
N_1/N_1 \approx \sqrt{1/N_1}$ (assuming Poisson statistics). Thus the
simulation yields the quantity $q = \exp(-\beta \Delta F_0)$ with the relative
error 
\begin{equation}
\label{eq:df1}
\Delta q/q = \exp(\beta \Delta F_0/2) \cdot \sqrt{1/N}. 
\end{equation}
In  the second setup, ($k$-step simulation), each simulation $i$ ($i =
1,\cdot,k$) gives a quantity $q_i = \exp(- \beta \Delta F_i)$ with the relative
error $\Delta q_i/q_i = \exp(\beta \Delta F_i/2) \cdot \sqrt{k/N}$. Ideally,
$\Delta F_i = \Delta F_0/k$ for all $i$, and the total relative error of $q =
\prod_i q_i$ accumulates to 
\begin{equation}
\label{eq:df2}
\Delta q/q = \exp(\beta \Delta F_0/2k) \cdot k/\sqrt{N}. 
\end{equation}
Comparing (\ref{eq:df1}) and (\ref{eq:df2}), one finds that a multi-step
strategy is favorable as soon as the free energy difference $\Delta F_0$ 
exceeds $4 \ln(2) \sim 2.77$ $k_B T$, and the optimal step size is 
characterized by a free energy difference of $ \beta \Delta F_0/k = 2 k_B T$. 

\subsection{Switching off the well potentials}
\label{sec:switch_wells}

Finally, in the last step, we gradually turn the well potentials to zero.
We evaluate the free energy difference $\Delta F_2$ between the intermediate
model $H'(\lambda=1,\varepsilon)$ and the target model
$H'(\lambda=1,\varepsilon=0) = H_0$ by a thermodynamic integration, using
\begin{equation}
\label{eq:TI}
F_0 - F'(1,\varepsilon) 
= - \int_{0}^{\varepsilon} \! {\rm d} \varepsilon'
\left\langle \frac{\partial H'(1,\varepsilon')}{\partial \varepsilon'} 
\right\rangle_{\varepsilon'} 
= - \int_{0}^{\varepsilon} \! {\rm d} \varepsilon'
\left\langle
\sum_i \Phi\Big(\frac{|\mathbf{r}_i - \mathbf{r}_i^{\rm ref}|}{r_{\rm cutoff}}\Big)
\right\rangle_{\varepsilon'}.
\end{equation}
Thus we need to sample the quantity $\langle \sum_i \Phi(|{\rm r}_i - {\rm
r}^{\rm ref}_i|/r_{\rm cutoff}) \rangle$ on an integration path between
$\varepsilon'=\varepsilon$ and $\varepsilon'=0$.  At this point, it becomes
clear why our potential wells must have a finite range $r_{\rm cutoff}$: If the
wells had infinite range, sampling $\langle \Phi \rangle$ in fluids would be
practically impossible in the limit $\varepsilon'=0$, where the mean-square
displacement of the particles diverges.

\begin{figure}[t]
\begin{center}
\includegraphics[width = .6 \columnwidth] {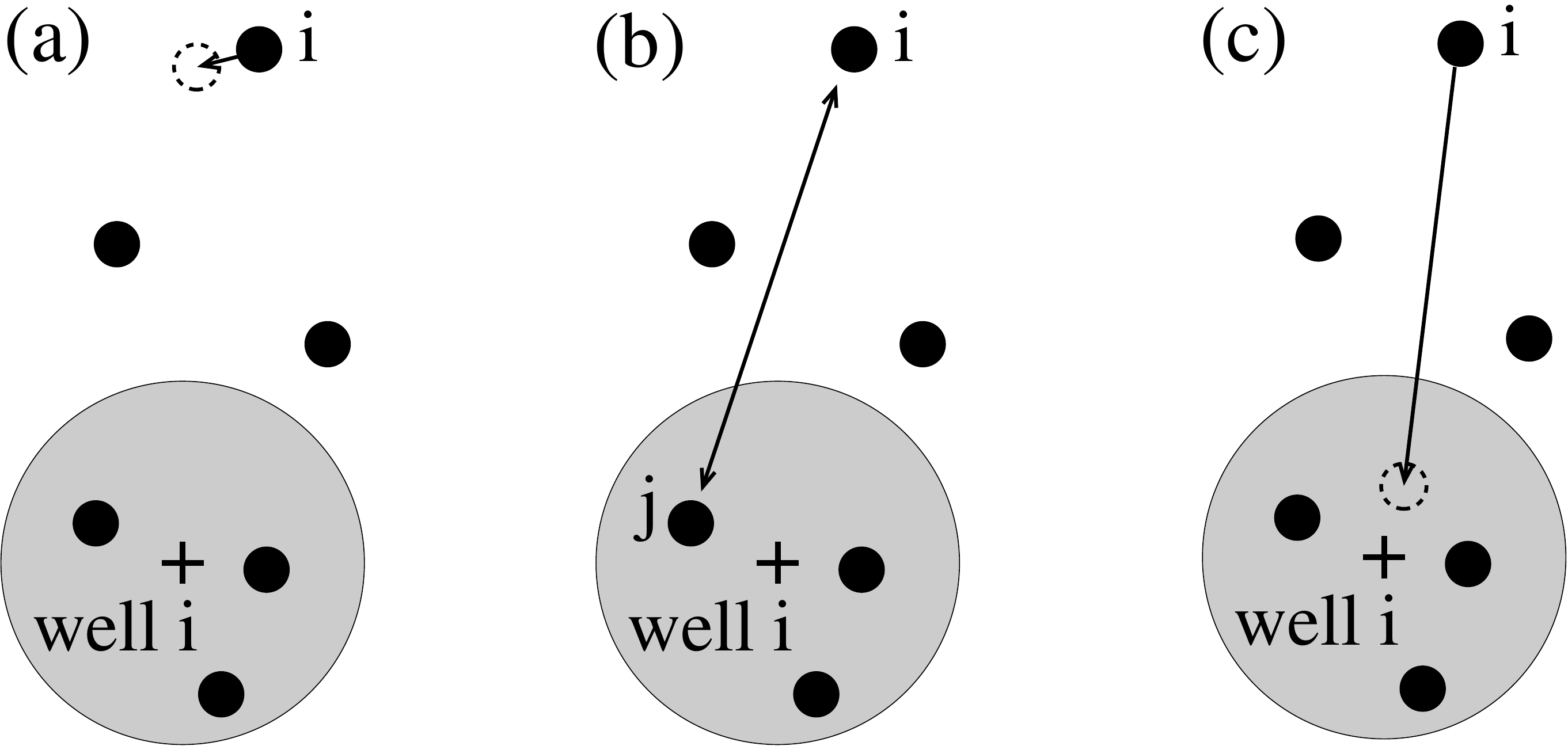} 
\vfill
\caption{Sketch of moves in our Monte Carlo algorithm.
(a) Simple particle displacements. (Could be replaced
{\em e.g.}, by short Molecular Dynamics runs.)
(b) Smart particle swaps.
(c) Smart particle relocations.
See text for explanation} \label{fig:moves}
\end{center}
\end{figure}

Setting a finite range for the reference potential, however, introduces a
problem: The particles need to find their respective wells of attraction. We
therefore introduce two Monte Carlo moves that help particles $i$ explore their
well $i$ (Fig.~\ref{fig:moves}). The first move (Fig.~\ref{fig:moves} b) swaps
particles in a smart way and works as follows:
\begin{itemize}
\item
  Pick a random particle $i$ and find the set of particles $\{n_i\}$
  that are within the attraction range of well $i$.
\item
  If particle $i \notin \{n_i\}$: pick a particle $j$ from $\{n_i\}$,
  and swap $i$ and $j$ with the probability 
    $\min\{1,\frac{n_i}{N}e^{-\Delta H'}\}.$
\item
  Otherwise: pick a particle $j$ from all particles  \\
  - if $j \notin \{n_i\}$: swap with probability
      $\min\{1,\frac{N}{n_i}e^{-\Delta H'}\}$. \\
  - if $j \in \{n_i\}$: swap with probability
      $\min\{1, e^{-\Delta H'}\}.$
\end{itemize}
Here $\Delta H'$ is the difference of the energies (according to the
intermediate model) of the old and new configuration. This algorithm promotes
particle swaps that bring particles close to their respective well and
nevertheless satisfies detailed balance.  

The second move (Fig.~\ref{fig:moves} c) relocates particles $i$ with a bias
towards the neighborhood of their well $i$:
\begin{itemize}
\item
  Pick a random particle $i$ (with position $\mathbf{r}_i$).
\item
  Choose a new position $\mathbf{r}'_i$ from a given 
  (biased) distribution $P_i(\mathbf{r}'_i) 
   = \exp(-W(|\mathbf{r}'_i - \mathbf{r}_i^{\rm ref}|))$. 
\item
 Relocate the particle from $\mathbf{r}$ to $\mathbf{r}'_i$ with probability
  $\min\{1,P(\mathbf{r}_i)/P(\mathbf{r}'_i) \: e^{-\Delta H'}\}$.
\end{itemize}
Obvious choices for $W(r)$ which we have tested are $W(r) = \varepsilon
\Phi(r/r_{\rm cutoff})$, or $W(r) = {\rm const.}$ for $r < r_{\rm cutoff}$.  As
long as the wells are weak (low $\varepsilon'$), the relocation moves have
little effect. At high $\varepsilon'$, they help to overcome trapped situations
where most particles are bound to a well, and a few cannot escape from a local
cage.

\begin{figure}[t]
\begin{center}
\includegraphics[width = 1. \columnwidth] {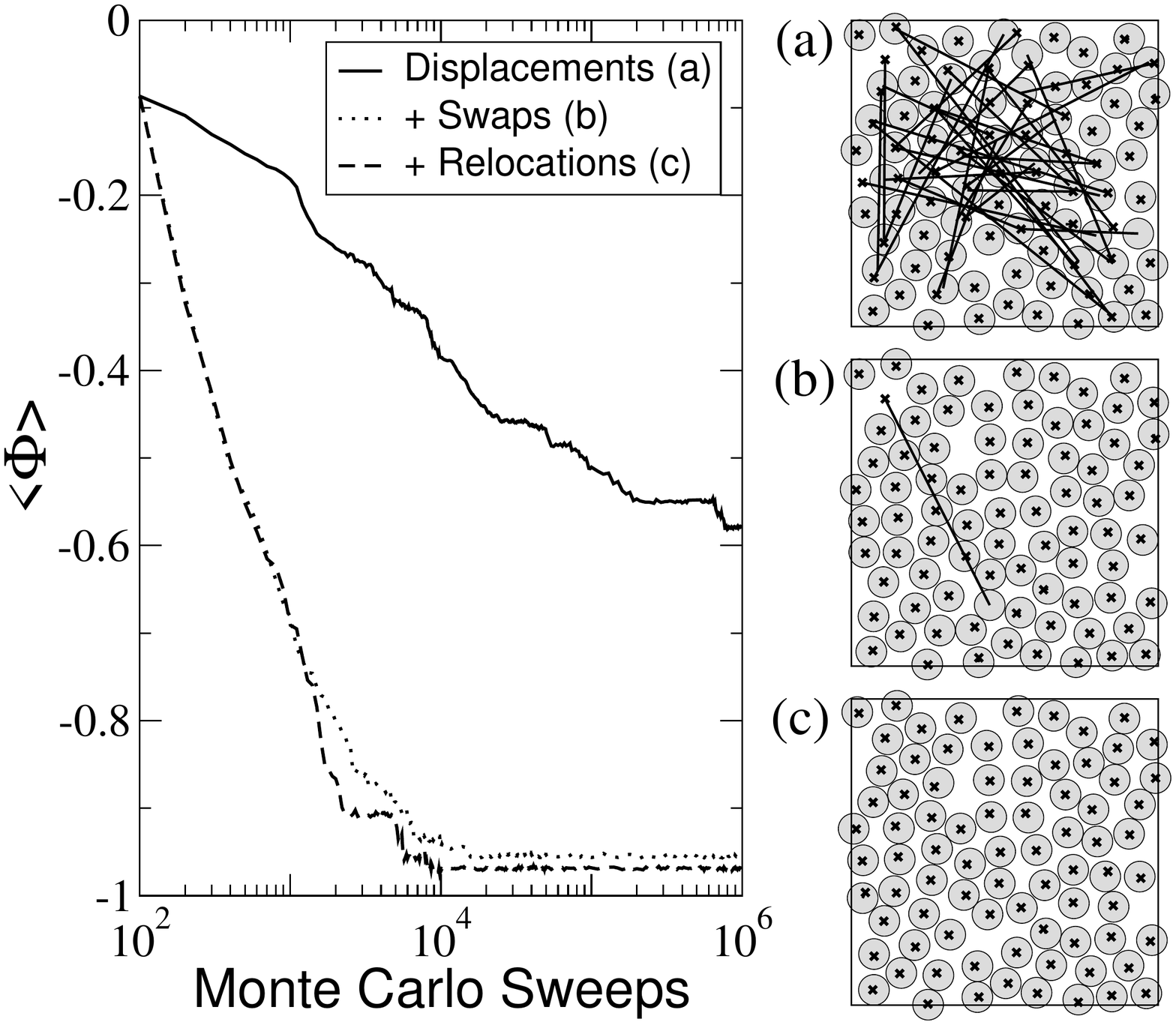} 
\vspace*{-0.1cm}
\caption{Illustration of the effect of the moves of
Fig.~\protect\ref{fig:moves} on the equilibration of a system of 80 hard disks
(diameter $D$) at a density $\rho = 0.8/D^2$, after switching on linear well
potentials with strength $\varepsilon = 50 k_B T$ ($r_{\rm cutoff} = 2 D$).  Swap
moves and relocation moves (one per bead) were attempted one per 100 MC sweeps.
Left: Evolution of $\langle \Phi \rangle$ in simulations that include different
moves as indicated.  Right: Corresponding final configurations.  Circles
indicate particle positions, crosses give well positions. Particles and their
respective wells are connected by straight lines. See text for more explanation.
From Ref.~\protect\cite{schilling09}.
} 
\label{fig:equil}
\vspace*{-0.5cm}
\end{center}
\end{figure}

The effect of the different moves is shown in Fig.~\ref{fig:equil} (taken from
Ref.~\cite{schilling09}). It shows the evolution of the observable $\langle
\Phi_i \rangle$, averaged over all particles $i$, in a two dimensional system
of hard disks (diameter $D$) at packing fraction $\eta = 0.63$, after
$\varepsilon$ had been raised from zero to $ \varepsilon = 50 k_B T$. The well
potential is linear (Eq.~(\ref{eq:phi_linear})) with range $r_{\rm cutoff} =
2D$.  In a Monte Carlo simulation that includes only random particle
displacements, the system is still far from equilibration after one million MC
sweeps (a).  Smart swap moves (one per 100 Monte Carlo sweeps) speed up the
equilibration by orders of magnitude, but the system gets trapped in a
configuration where one particle remains separated from its well (b). This
problem is solved by including smart relocation moves (c).

\section{Extensions of the method}
\label{sec:extensions}

The algorithm described above can easily be generalized to molecular fluids:
one must simply swap molecules instead of particles (atoms).  To generalize it
to free enthalpy calculations (systems at constant pressure), we employ a
reference system that is defined in terms of scalable coordinates, {\em i.e.},
the reference sites are rescaled along with the particle coordinates if the
volume of the system changes. Furthermore, the volume $V$ of the system is
pinned to the reference volume by an additional term $\varepsilon
(V-V_{\rm ref})^2$ in the reference Hamiltonian. We can then follow the strategy
described above, with two modifications: First, the derivative of the free
energy in Equation (\ref{eq:TI}) has an additional contribution
\begin{equation}
\label{eq:dfde_enthalpy}
\frac{\partial F'}{\partial \varepsilon} = 
\big\langle \frac{\partial H'}{\partial \varepsilon} \big\rangle = 
\big\langle \sum_i \Phi \Big(
\frac{|{\rm r}_i - {\rm r}^{\rm ref}_i|}{r_{\rm cutoff}} \Big) \big \rangle
+ \big \langle (V - V_{\rm ref})^2 \big \rangle.
\end{equation}
Second, the expression for the free enthalpy of the reference system changes. 
In practice, the fluctuations of the volume $V$ become negligible already at 
moderate $\varepsilon$, ($\varepsilon > 10 k_B T$ in our case),
such that the reference free energy can be written as
\begin{equation}
\label{eq:ref_FreeEnthalpy}
  \frac{\beta \: G_{\rm ref}(\varepsilon)}{N} 
    = \frac{P V_{\rm ref}}{N} - 1 + \ln\left(\frac{N}{V_{\rm ref}}\right) 
      - \ln\left(1 +\frac{V_0}{V_{\rm ref}} g_{\Phi}(\beta \varepsilon)\right)
\end{equation}
to a very good approximation.

\section{Results}
\label{sec:results}

\subsection{Transition to the reference state: Effect of well shape and well range}

\begin{figure}[t]
\begin{center}
\includegraphics[width = .7 \columnwidth] {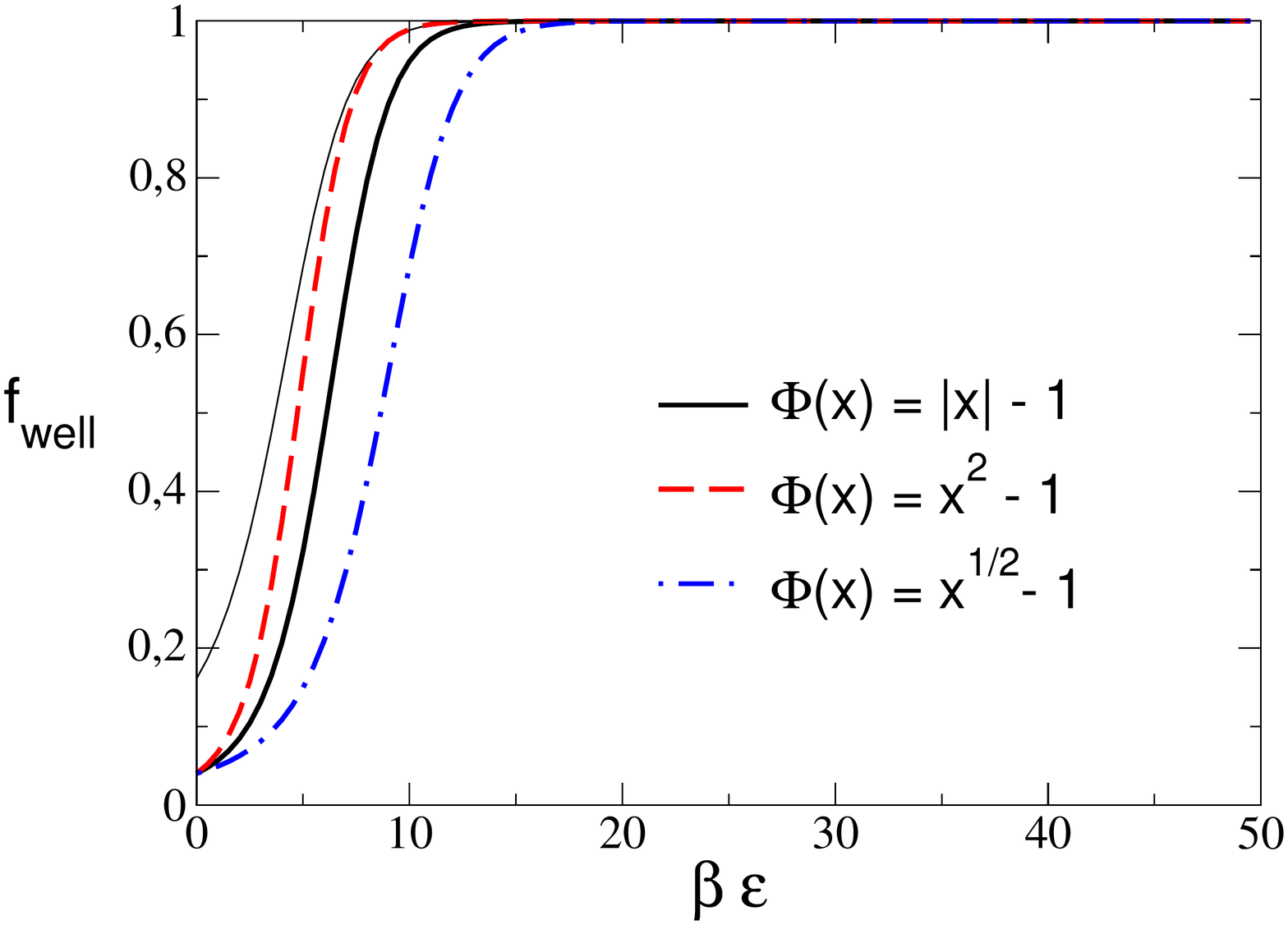} 

\vspace*{-1cm}
\includegraphics[width = .7 \columnwidth] {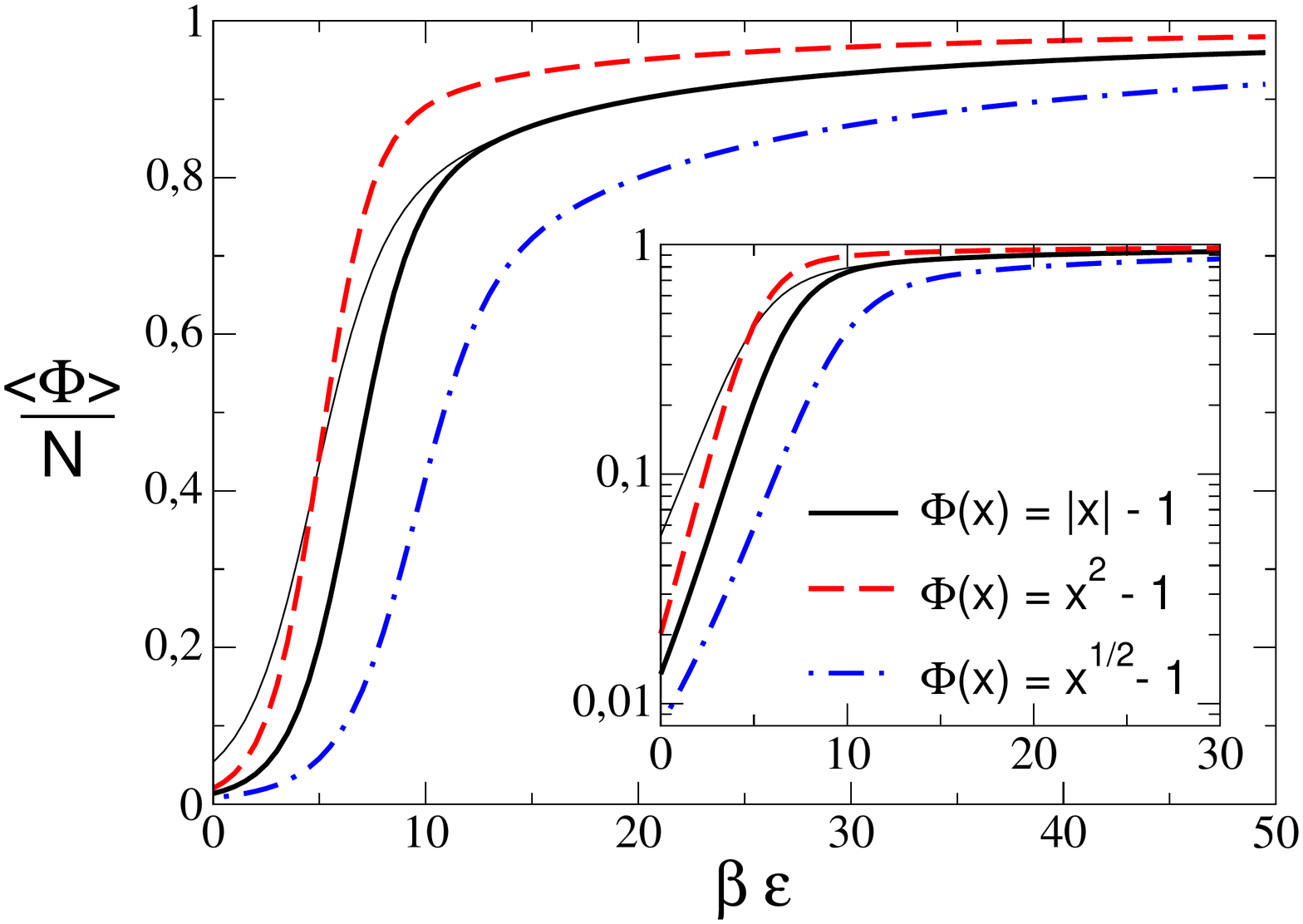} 

\vspace*{-1cm}
\includegraphics[width = .7 \columnwidth] {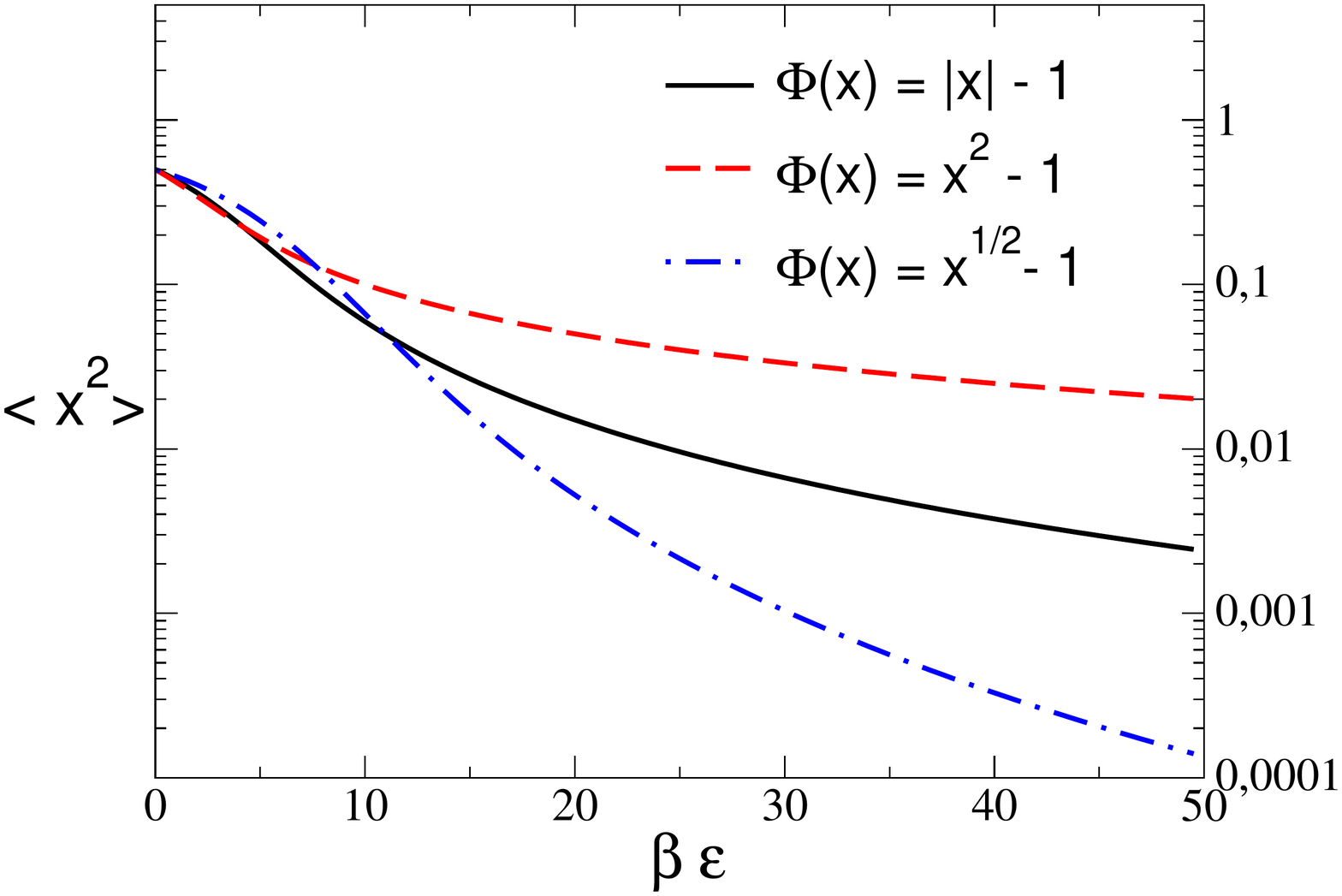} 
\vfill
\caption{Effect of well shape $\Phi(x)$ on the behavior of different quantities
as a function of well strength $\varepsilon$ in the two dimensional ideal gas. 
Top: Fraction $f_{\rm well}$ of particles inside the well.
Middle: Free energy derivative per particle $\langle \Phi \rangle/N.
=-\partial F /\partial \varepsilon \cdot N^{-1}$ .
Bottom: Mean-square displacement of those particles that are within the well range.
Dashed line corresponds to a quadratic well, thick solid line to a linear well, and
dot-dashed line to a square root well with well range $\sigma$. The thin solid
line shows the behavior of the same quantities for a linear well with range 
$2 \sigma$ for comparison. The volume is $V = 77.94 \sigma^2$ (Here $\sigma$
is an arbitrary length scale).
}
\label{fig:well_shapes}
\end{center}
\end{figure}

Before presenting our first real application examples, we will discuss a few
properties of the algorithm at the example of the two-dimensional ideal gas.
Figure \ref{fig:well_shapes} illustrates the influence of the shape and range
of the well potential.  Three well shapes are compared, the (most popular)
harmonic well, the linear well (Equation (\ref{eq:phi_linear}), and a ''square
root'' well with the shape $\Phi(x) = \sqrt{x}-1$. The top panel in Figure
\ref{fig:well_shapes} demonstrates that of these three potentials, the harmonic
well is most effective in trapping the particles for small and moderate
$\varepsilon$. All three potentials however bind almost 100 \% of all particles
for $\varepsilon > 20 k_B T$. At higher $\varepsilon$, the linear and the
square root potential localize the particles much better (lowest panel in
Figure \ref{fig:well_shapes}), which makes it more easy to switch on
interactions (step \ref{sec:switch_inter}) in the case of interacting
particles.  The middle panel in Figure \ref{fig:well_shapes} shows the averaged
well energy per particle $\langle \Phi \rangle/N$, {\em i.e.}, the quantity
that needs to be sampled for the thermodynamic interactions. It is remarkable
that $\langle \Phi \rangle/N$ is still noticeably different from its limiting
value ($\langle \Phi \rangle/N = 1$) even for $\varepsilon$-values as high as
$\varepsilon = 50 k_B T$. Despite the fact that it localizes particles most
efficiently, the square well potential reaches the limiting value more slowly
than the other potentials. In the following studies, we will mostly settle on
using linear well potentials.

\clearpage

\subsection{Transition to the reference state: Effect of packing fraction}

Next we study the effect of particle interactions. To this end, we introduce
hard disk interactions in the system of Figure \ref{fig:well_shapes} (using the
linear well potential with range $r_{\rm cutoff} = 2 \sigma$).  By adjusting
the disk radius, we vary the packing fraction $\eta$ between $\eta = 0$ and
$\eta = 0.9$, which is close to the maximum value $\eta_{\rm max} = 0.9069$. At
the two higher packing fractions, $\eta = 0.7$ and $\eta = 0.9$, the stable
state is an ordered crystal structure.  At $\eta=0.7$, we also show for
comparison the curve obtained with a (metastable) disordered reference state. 

\begin{figure}[t]
\begin{center}
\includegraphics[width = 1.0 \columnwidth] {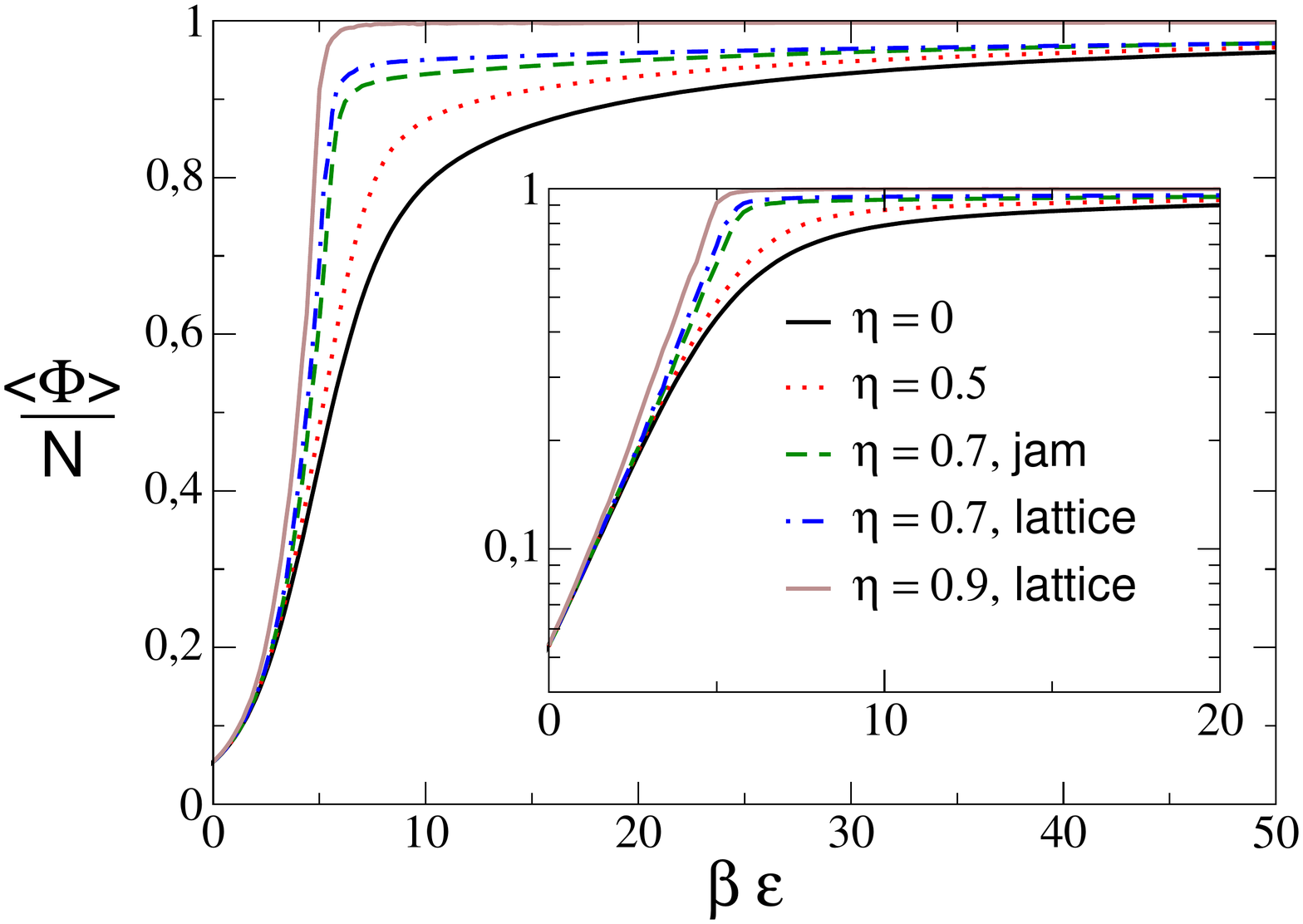} 
\vfill
\caption{ Free energy derivative per particle $\langle \Phi \rangle/N
=-\partial F /\partial \varepsilon \cdot N^{-1}$ in a system of hard disk
particles with different diameters, {\em i.e.}, different packing fractions
$\eta$. The shape of the well potential is linear with cutoff $r_{\rm cutoff} =
2 \sigma$.  The other parameters are the same as in
Fig.~\protect\ref{fig:well_shapes}.
}
\label{fig:phi_eta}
\end{center}
\end{figure}

With increasing packing fraction, the particles settle more rapidly in their
respective wells. This also happens faster for the ordered system than for the
jammed system. To check for hysteresis effects, the curves were sampled in both
directions, with {\em i.e.}, increasing and decreasing $\varepsilon$. We did
not encounter any signs of a first order transition. At the highest packing
fraction $\eta=0.9$, the equilibration times were found to become very large for
small $\varepsilon$.  The reason is that in the absence of wells, the center of
gravity of the whole crystalline system diffuses around very slowly. This
problem is already well known from the 'Einstein crystal' method and has led to
the development of the 'Einstein molecule' method, where the reference system
is defined in terms of relative coordinates rather than absolute coordinates
\cite{vega07}. The same idea can also be used here.

\subsection{Application: Hard sphere fluids}

After these more technical discussions, we turn to presenting real
applications. As a first application example, we have determined the absolute
free energies of systems of 256 hard spheres with diameter $D$ at the densities
$N/V = 0.25 D^{-3}, 0.5 D^{-3}$, and $0.75 D^{-3}$. The results can be compared
with those obtained by integration of the Carnahan-Starling
equation-of-state\cite{carnahan69}, which describes the behavior of three
dimensional hard sphere fluids very accurately. To obtain the free energies, we
have used 50 values of $\varepsilon$ and $6 \times 10^5$ Monte Carlo sweeps for
each data point for the two lower densities, and 200 values of $\varepsilon$
times one million Monte Carlo sweeps for the density $0.75 D^{-3}$. The results
are given in Table \ref{tab:3dspheres}.  Within the error, all results agree
well with the theoretical value from the Carnahan-Starling equation.

For the density $0.5 D^{-3}$ (corresponding to a liquid state), we compared
three different situations: (a) linear well $\Phi$ and liquid reference state,
(b) linear well $\Phi$ and crystalline reference state, (b) harmonic well
$\Phi$ and liquid reference state. All variations gave the same results.  In
case (b), we did not see a hysteresis on increasing/decreasing $\varepsilon$ --
apparently, the trapping of particles in a crystalline array of wells is not
associated with a phase transition at this density. We expect that this will be
different closer to the liquid/solid transition. Nevertheless, we can conclude
that our method is robust and may work if the reference configuration is not
really representative for the target structure. Comparing the calculation for
linear and harmonic wells, we find that the result also does not depend on the
shape of the well. However, for more accurate potential, the linear potential
seems to be more useful, because the particles get trapped more efficiently.

\begin{table}[ht]
  \caption{\label{tab:3dspheres}Results for the free energy of hard spheres
    (from Ref. \protect\cite{schilling09}). ${F/N}_{\rm CS}$ is the value according 
    to the Carnahan-Starling equation-of-state \cite{carnahan69}. 
    a) linear potential $\Phi$, liquid reference state. 
    b) linear $\Phi$, hcp reference state. c) harmonic $\Phi$, liquid
    reference state.}
\begin{center}
    \begin{tabular}{ccc}
      $N/V$ & $F/N$ & $(F/N)_{\rm CS}$ \\
      \hline
      $0.25$ & $0.620 \pm 0.002$ & $0.625$ \\
      $0.5^{a)}$ & $1.541 \pm 0.002$ &$1.544$ \\
      $0.5^{b)}$ & $1.540 \pm 0.002$ &$1.544$\\
      $0.5^{c)}$ & $1.549 \pm 0.002$ &$1.544$\\
      $0.75$     & $3.009 \pm 0.005$ &$3.005$ \\
    \end{tabular}
\end{center}
\end{table}

\subsection{Free energy of a vacancy}

\begin{figure}[t]
\begin{center}
\includegraphics[width = 1.0 \columnwidth] {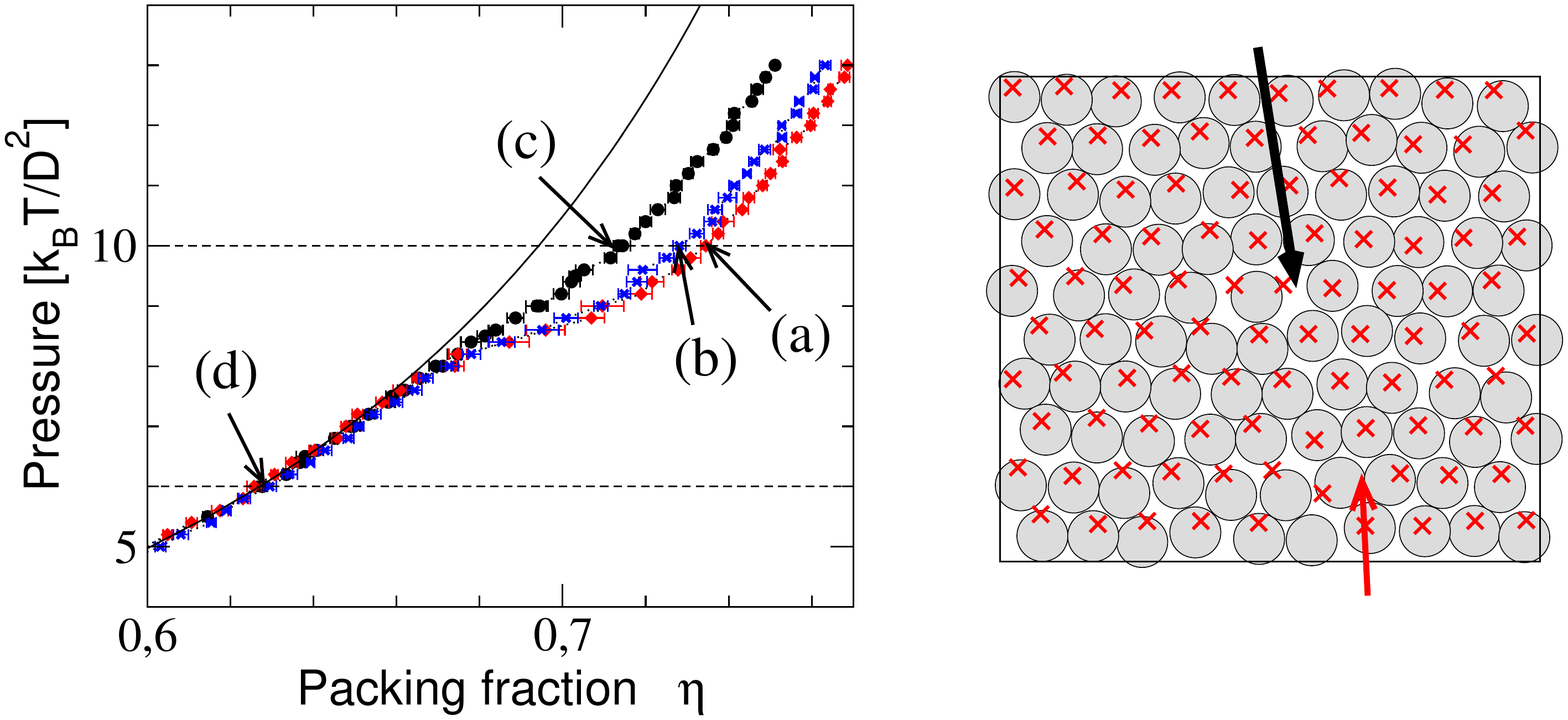}  \\
\vspace*{-3cm}
\includegraphics[width = 1.0 \columnwidth] {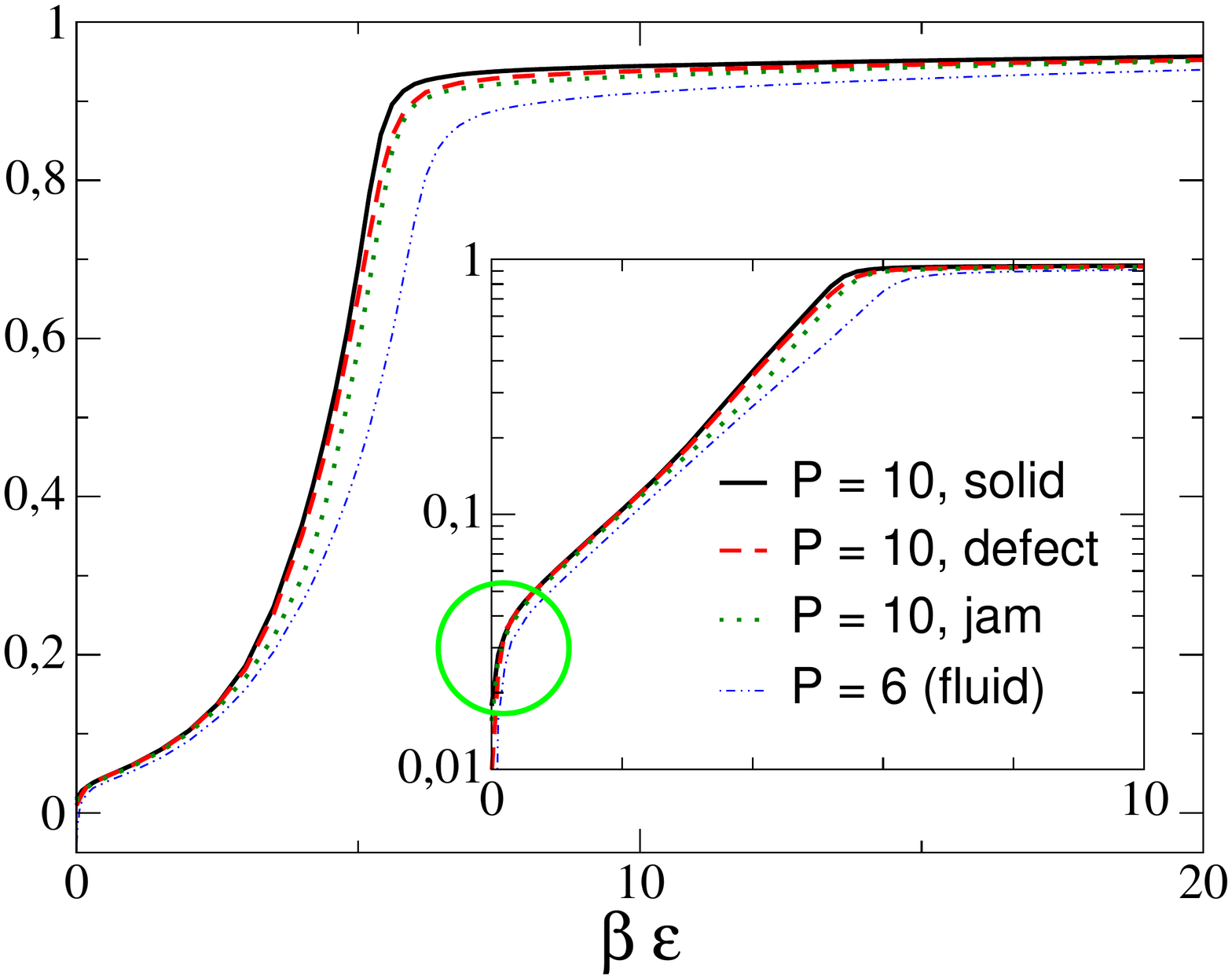} 
\vfill
\caption{Free enthalpy calculations of dense hard-disk systems
(diameter $D$).
Top left: Pressure vs. packing fraction as obtained from constant pressure
simulations at $\varepsilon = 0$. (a) expanded from an ordered
solid phase, (b) expanded from a solid phase with one vacancy,
(c) compressed from the fluid phase. The solid line shows
the theoretical estimate \cite{helfand60} 
$P = \hat{\rho}/(1 - \hat{\rho} \cdot \pi/4)^2$ with
$\hat{\rho} = (N+1)/\langle V \rangle$.
Top right: Configuration snapshots with one vacancy at the 
beginning (crosses) and the end (circles) of a Monte Carlo
run at $\varepsilon = 0$. The arrows mark the position of the
defect in the two cases, illustrating its high mobility.
Bottom: Derivative $- \partial F/\partial \varepsilon$
of the free energy along the path of thermodynamic integration
for the three dense systems (a),(b), and (c) at pressure
$P=10 k_B T/D^2 $, and for a fluid system at $P=6 k_B T/D^2$.
The inset shows the same data in a semi-logarithmic scale.
The circle highlights the area where the volume fluctuations
contribute noticeably to $\partial F/\partial \varepsilon$.
}
\label{fig:defect}
\end{center}
\end{figure}

Our last example shows an application of the method to a dense disordered
system, where the dynamics is driven by cooperative processes.  We have studied
hard disks (two dimensions, diameter $D$) up to densities where the equilibrium
state is solid, and enforced a vacancy defect by taking one particle out of an
otherwise ordered configuration. To calculate the free energy of the vacancy,
we must compare the {\em enthalpy} of this system with that of an ordered
system at the same pressure. Thus simulations were carried out a constant
pressure (see Section \ref{sec:extensions}) in a rectangular simulation box of
varying area and fixed side ratio $1:\sqrt{3/4}$.  The vacancy is then stable,
but highly mobile (Figure \ref{fig:defect}, top right).

\clearpage

We compare four different structures (Figure \ref{fig:defect}): An ordered
solid (a), an ordered solid with a vacancy (b), and for comparison also a
metastable disordered jammed state (c), which was obtained by compressing the
system from the fluid state, and a fluid state (d) at lower pressure.  Figure
\ref{fig:defect} (top left) shows the corresponding pressure-density curves of
the target system ($\varepsilon=0$). Free enthalpy calculations were carried
out at $P=10 k_B T/D^2$ for the three cases (a-c), and at $P=6 k_B T /D^2$ for
the case (d). The resulting free enthalpies can be related to the chemical
potential $\mu$ by virtue of the thermodynamic relation $G = \mu N$. 
Figure \ref{fig:defect} shows the corresponding evolution of the integrands
$\langle \partial H'/\partial \varepsilon \rangle$ of the thermodynamic
integration (Equation (\ref{eq:dfde_enthalpy}) up to $\varepsilon = 20$.
(in total, $\varepsilon$ was varied between 0 and $50 k_B T$). The volume
fluctuations contribute only for very small $\varepsilon$. The curves for
the jam, solid, and defect state are slightly different, reflecting the
differences in the free enthalpies of the different states.

At $P=6 k_B T/D^2$, the free enthalpy calculation yields the chemical potential
$\mu = 8.997 \pm 0.002 k_B T$, which is in good agreement with the theoretical
estimate of Helfand {\em et al} \cite{helfand60} $\mu = 9.047 k_B T$. At $P=10 k_B
T/D^2$, we find $\mu_{\rm solid} = 13.617 \pm 0.002 k_B T$ in the solid state,
and $\mu_{\rm jam} = 13.675 \pm 0.002 k_B T $ in the jammed state, which
establishes that the solid state is indeed the stable phase. For the system
with one defect, we obtained the total enthalpy $G_{\rm defect} = 1361.7 \pm
0.2 k_B T$.  This result can be used to estimate the core free energy of the
vacancy $\mu_c = G_{\rm defect}- \mu_{\rm solid} N  + k_B T \ln (N)= 7.1 \pm
0.3 k_B T$, which corresponds to a relative vacancy frequency of roughly
$10^{-3}$.  (For comparison, the frequency of vacancies at liquid/solid
coexistence in three dimensions is roughly $10^{-4}$, according to Pronk {\em
et al}\cite{pronk01}).  Since finite size effects stabilize the defect, $\mu_c$
is probably largely overestimated, hence the value given above has to be
interpreted as an upper bound. More detailed studies shall be carried out in
the future.  Here, the example mainly serves to illustrate the use of our
approach in situations where free energies are difficult to access with other
methods.

\section{Conclusions}

\label{sec:summary}

In summary, we propose a new method to compute absolute free energies for
fluids, liquids, and disordered structures.  The method generalizes the
Einstein crystal method of Frenkel and Ladd \cite{frenkel84}, but can be
applied much more widely. We anticipate that the method will be useful for
studies of complex fluids and biological systems, {\em e.g.}, proteins in
solution. This remains to be tested. The method has recently been published
\cite{schilling09}, and a few application examples have been given. We hope
that the detailed discussion in the present proceedings paper will motivate
researchers to try and apply the method to many other systems and further
explore its potential.





\bibliographystyle{elsarticle-num}
\bibliography{<your-bib-database>}



\end{document}